\documentstyle[epsfig,aps]{revtex} 
\begin{document} 
   
\def\bea{\begin{eqnarray}}
\def\eea{\end{eqnarray}}
\def\beas{\begin{eqnarray*}}
\def\eeas{\end{eqnarray*}}
\def\nn{\nonumber}
\def\ni{\noindent}
\def\G{\Gamma}
\def\d{\delta}
\def\l{\lambda}
\def\g{\gamma}
\def\m{\mu}
\def\n{\nu}
\def\s{\sigma}
\def\b{\beta}
\def\a{\alpha}
\def\f{\phi}
\def\fh{\hat{\phi}}
\def\y{\psi}
\def\z{\zeta}
\def\p{\pi}
\def\e{\epsilon}
\def\ve{\varepsilon}
\def\cl{{\cal L}}
\def\cv{{\cal V}}
\def\cz{{\cal Z}}
\def\pl{\partial}
\def\ov{\over}
\def\~{\tilde}
\def\rar{\rightarrow}
\def\lar{\leftarrow}
\def\lrar{\leftrightarrow}
\def\rra{\longrightarrow}
\def\lla{\longleftarrow}
\def\8{\infty}

%\twocolumn[\hsize\textwidth\columnwidth\hsize\csname@twocolumnfalse%   
%\endcsname
\title{
Renormalization Conditions and the Effective Potential
of the Massless $\phi^4$ Theory}    
\author{J.~-M. Chung\footnote        
  {Electronic address: jmchung@photon.kyunghee.ac.kr}         
  ~and B.~K. Chung}     
\address{ Research Institute for Basic Sciences\\         
and Department of Physics,\\ Kyung Hee University, Seoul  130-701, Korea}    
    
\date{\today}    
\maketitle    
\draft    
\begin{abstract}         
We point out that there is a missing portion in the two-loop effective 
potential of the massless $O(N)$ $\phi^4$ theory obtained by Jackiw in
his classic paper, Phys. Rev. D 9, 1686 (1974). 
\end{abstract}    
         
%]
%%%%%%%%%%%%%%%%%%%%%%%%%%%%%%%%%%%%%%%%%%%%%%%%%%%%%%%%%%%%%%%%%%%%%%%%%%%%% 
\section{Introduction}   
%%%%%%%%%%%%%%%%%%%%%%%%%%%%%%%%%%%%%%%%%%%%%%%%%%%%%%%%%%%%%%%%%%%%%%%%%%%%% 
The effective potential in quantum field theory plays a crucial role 
in connection with the problem of the spontaneous symmetry breaking.
In this field there are three classic papers \cite{cw,jk,iim}.
Coleman and Weinberg\cite{cw} were the first ones to calculate  
the higher-order effective potential of a scalar field at one loop  
level by summing up an infinite number of Feynman graphs.  
Jackiw\cite{jk} has used the Feynman path-integral method to obtain 
a simple formula for the effective potential. He has succeeded in 
representing each loop order containing 
an infinite set of conventional Feynman graphs by finite number of graphs 
using this algebraic method which can formally be extended to the arbitrary 
higher-loop order. 
In Ref.~\cite{iim} the functional integral is explicitly evaluated 
using the steepest descent method at two-loop level. Higher-loop  
calculations with this method are very difficult.

The purpose of this paper is to show that there is a missing portion in 
the two-loop effective potential of the massless $O(N)$ $\f^4$ theory 
obtained by Jackiw \cite{jk}.
In this paper we employ the dimensional regularization method \cite{tv}
instead of the cutoff regularization method used in Ref.~\cite{jk} and
for the sake of brevity we confine ourselves to the case of 
single component theory ($N=1$). 

%%%%%%%%%%%%%%%%%%%%%%%%%%%%%%%%%%%%%%%%%%%%%%%%%%%%%%%%%%%%%%%%%%%%%%%%%%%% 
\section{Review of the Calculation}       
%%%%%%%%%%%%%%%%%%%%%%%%%%%%%%%%%%%%%%%%%%%%%%%%%%%%%%%%%%%%%%%%%%%%%%%%%%%% 
The Lagrangian for a theory of a self-interacting spinless field $\f$ is given 
as       
\bea       
{\cal L}(\f(x))&=&{1+\d Z\ov 2}\pl_\m\f\pl^\m\f 
-{m^2+\d m^2\ov 2}\f^2-{\l+\d\l\ov 4!}\f^4\;, \label{lg}       
\eea     
where the quantities $\f$, $m$,      
and $\l$ are the renormalized field, the renormalized mass, and  
the renormalized coupling constant respectively, whereas $\d Z$, $\d m^2$,  
and $\d\l$ are corresponding (infinite) counterterm constants.  
We will confine ourselves     
to the massless theory ($m=0$).  
The effective potential is most suitably defined, when the    
effective action $(\G[\f_{\rm cl}])$, being the generating functional of  
the one-particle-irreducible (1PI) Green's functions  
($\G^{(n)}(x_1,...,x_n)$),   
is expressed in the following local form (the so-called derivative    
expansion):   
\bea   
\G[\f_{\rm cl}]&=&\int\!d^4x\biggl[-\cv(\f_{\rm cl}(x)) 
+{1\ov 2}\cz(\f_{\rm cl}(x))\pl_\m\f_{\rm cl}(x)
\pl^\m\f_{\rm cl}(x)+\cdots~\biggr]\;,  \label{loc}   
\eea   
where $\f_{\rm cl}(x)$ is the vacuum expectation    
value of the field operator $\f(x)$ in the presence of an external source.    
By setting $\f_{\rm cl}(x)$ in $\cv(\f_{\rm cl}(x))$ to be a constant    
field $\fh$, we obtain the effective potential $V_{\rm eff}(\fh)$   
\bea   
V_{\rm eff}(\fh)\equiv \cv(\f_{\rm cl}(x))|_{\f_{\rm cl}(x)=\fh}\;.\label{ve}   
\eea   
   
Following the field-shift method of Jackiw \cite{jk}   
for the calculation of the effective potential, we first obtain the shifted     
Lagrangian with the constant field configuration $\fh$     
\bea       
{\cal L}(\fh;\f(x))&=&{1+\d Z\ov 2}\pl_\m\f\pl^\m\f 
-{1\ov 2}\biggl(\d m^2+{\l+\d\l\ov 2}\fh^2\biggr)\f^2\nn\\     
&-&{\l+\d\l\ov 6}\fh\f^3-{\l+\d\l\ov 4!}\f^4\;. \label{slg}       
\eea     
The Feynman rules for this shifted Lagrangian are given in Fig.~1.     
%%%%%%%%%%%%%%%%%%%%%%%%%%%%%%%%%%%%%%%%%%%%%%%%%%%%%%%%%%%%%%%%%%%%%%%%%%%%% 
%  FIGURE                                                                   %
%%%%%%%%%%%%%%%%%%%%%%%%%%%%%%%%%%%%%%%%%%%%%%%%%%%%%%%%%%%%%%%%%%%%%%%%%%%%%
\begin{figure}
  {\unitlength1cm 
   \epsfig{file=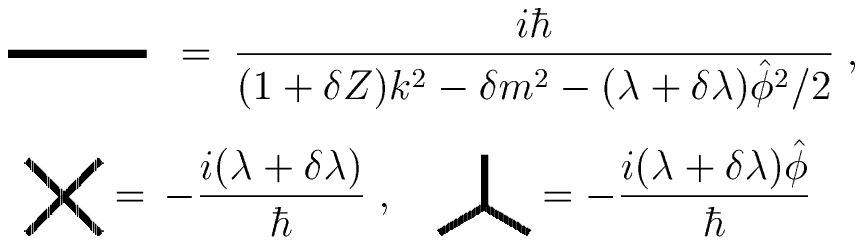, 
   bbllx=55pt,bblly=0pt,bburx=612pt,bbury=650pt, 
      rheight=3.5cm, rwidth=10cm,  clip=,angle=0} }   
\caption{Feynman rules of the shifted Lagrangian, Eq.~(4).}   
\end{figure}

Without introducing any new loop-expansion parameter, which is eventually  
set to be unity, we will use $\hbar$ as a loop-counting parameter \cite{nb}.  
This is the reason why we have kept all the traces of $\hbar$'s in the  
Feynman rules above in spite of  
our employment of the usual ``God-given'' units, $\hbar=c=1$.  
In addition to the above Feynman rules, Fig.~1, which are     
used in constructing two- and higher-loop vacuum diagrams, we need another      
rule (Fig.~2) solely for a one-loop vacuum diagram which is dealt with      
separately in Jackiw's derivation of his prescription and is essentially      
the same as that of Coleman and Weinberg \cite{cw} from the outset.
%%%%%%%%%%%%%%%%%%%%%%%%%%%%%%%%%%%%%%%%%%%%%%%%%%%%%%%%%%%%%%%%%%%%%%%%%%%%% 
%  FIGURE                                                                   %
%%%%%%%%%%%%%%%%%%%%%%%%%%%%%%%%%%%%%%%%%%%%%%%%%%%%%%%%%%%%%%%%%%%%%%%%%%%%% 
\begin{figure}[h] 
  {\unitlength1cm 
   \epsfig{file=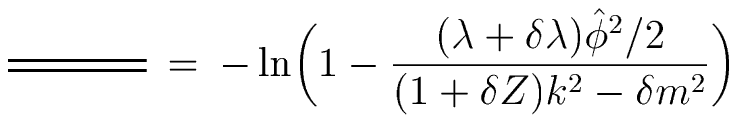, 
   bbllx=63pt,bblly=0pt,bburx=612pt,bbury=638pt, 
      rheight=1.5cm, rwidth=10cm,  clip=,angle=0} }   
\caption{Feynman rule for one-loop vacuum diagram.} 
\end{figure} 
   
Using the rules, Fig.~1 and Fig.~2, and including the terms    
of zero-loop order, we arrive at the formal expression of the effective     
potential up to two-loop order:   
\bea     
V_{\rm eff}(\fh)&=&\biggl[{\d m^2\ov 2}\fh^2     
+{\l+\d\l\ov 4!}\fh^4\biggr]+\Bigl[\mbox{Diag.~1}\Bigr]     
+\Bigl[\mbox{Diag.~2}\Bigr]+\Bigl[\mbox{Diag.~3}\Bigr]\;.\label{onn}     
\eea 
The last three (bracketed-) terms on the right-hand side in the 
above equation appear in Fig.~3.

%%%%%%%%%%%%%%%%%%%%%%%%%%%%%%%%%%%%%%%%%%%%%%%%%%%%%%%%%%%%%%%%%%%%%%%%%%%%% 
%  FIGURES                                                                  %
%%%%%%%%%%%%%%%%%%%%%%%%%%%%%%%%%%%%%%%%%%%%%%%%%%%%%%%%%%%%%%%%%%%%%%%%%%%%% 
\begin{figure} 
  {\unitlength1cm 
   \epsfig{file=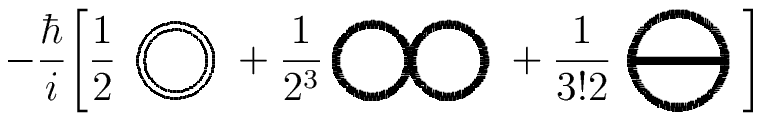, 
   bbllx=55pt,bblly=0pt,bburx=612pt,bbury=660pt, 
      rheight=2.5cm, rwidth=10cm,  clip=,angle=0} }   
\caption{[Diag.~1]+\,[Diag.~2]+\,[Diag.~3].} 
\end{figure} 

For the purposes of renormalization we first expand the counterterm  
constants in power series, beginning with order $\hbar$:  
\beas     
\d m^2&=&\hbar\d m_1^2+\hbar^2\d m_2^2+\cdots\;,\nn\\  
\d\l&=&\hbar\d \l_1+\hbar^2\d\l_2+\cdots\;,\nn\\  
\d Z&=&\hbar\d Z_1+\hbar^2\d Z_2+\cdots\;.       
\eeas     
In what follows we will use the following notation for the effective potential 
up to the $L$-loop order: 
\beas 
V_{\rm eff}^{[L]}(\fh)=\sum_{i=0}^L \hbar^i  
V_{\rm eff}^{(i)}(\fh)\;. 
\eeas 
The zero-loop part of the effective potential is given as 
\bea 
V_{\rm eff}^{(0)}(\fh)={\l\ov 4!}\fh^4\;. \label{ve0}
\eea 
The one-loop part of the effective potential is readily obtained as 
\bea 
V_{\rm eff}^{(1)}(\fh)&=&{\d m_1^2\ov 2}\fh^2+{\d \l_1\ov 4!} 
\fh^4 -{\l^2\fh^4\ov 8(4\p)^2\e}
+{\l^2\fh^4\ov (4\p)^2}\biggl[-{3\ov 32}+{\g\ov 16}+{1\ov 16}   
\ln\biggl({\l\fh^2/2\ov 4\p M^2}\biggr)\biggr]\;,\label{ve1} 
\eea 
where $\g$ is the usual Euler constant and $M$ is an arbitrary constant
with mass dimension.
The $\e$ poles in this equation are readily cancelled out by choosing 
the counterterm constants $\d m_1^2$ and $\d\l_1$ as follows: 
\bea     
\d m_1^2=a_1\;,~~~~~\d\l_1={3\l^2\ov (4\p)^2\e}+b_1\;, 
\label{11}
\eea 
where $a_1$ and $b_1$ are unspecified but finite constants at this stage.  
One may put $a_1$ (and $a_2$ below) to be zero from the beginning 
because the theory is massless. In our dimensional regularization scheme 
the pole part of $\d m_1^2$ vanishes, but this is not the case in the cutoff  
regularization method. 
Besides $\d m_1^2$ and $\d\l_1$, there is another counterterm constant.
It is $\d Z_1$. In Jackiw's calculation, $\d Z_1$ is set to be zero.
This is matched to the standard condition for the defining the scale of the
field
\bea
\cz|_{\fh=0}=1\;.\label{wzc}
\eea
In the massless theory, however, the above 
condition is afflicted by the infrared singularity, as remarked by
Coleman and Weinberg \cite{cw}. (In fact, this singularity cannot be seen
in $\cz^{(1)}$, the one-loop order contribution to $\cz$. The infrared
singularity appears for the first time in the two-loop order \cite{iz}.)

Now let us determine $\d Z_1$ so as to meet the following modified 
condition which avoids the infrared singularity:
\bea
\cz|_{\fh^2=M^2}=1\;.
\eea
To this end, we use the following relation \cite{gkf}
\bea
\cz|_{\fh^2=M^2}=
{\pl\~{\G}^{(2)}_{\fh}(p^2)\ov \pl p^2}\bigg|_{p^2=0,\,\fh^2=M^2}\;,
\label{mc}
\eea 
where $\~{\G}^{(2)}_{\fh}(p^2)$ is the (momentum-conserving) 1PI two-point 
Green's function in the shifted theory. The right-hand side of Eq.~(\ref{mc})
is calculated as 
\beas
1+{\hbar\l\ov 6(4\p)^2}+\hbar\d Z_1\;,
\eeas
from which we find
\bea
~~~~~~~\d Z_1=-{\l\ov 6(4\p)^2}\equiv c_1\;.\label{c1}
\eea
Note that this wave function renormalization constant $\d Z_1$ is free of
$\e$ singularity. But in a higher-loop order
the wave function renormalization constant $\d Z_n$ may have the 
$\e$ singularity.

The two-loop part of the effective potential is obtained as  
\bea 
V_{\rm eff}^{(2)}(\fh)&=&{\d m_2^2\ov 2}\fh^2+{\d \l_2\ov 4!}\fh^4     
-{a_1\l\fh^2\ov 2(4\p)^2\e}
-{3\l^3\fh^4\ov     
8(4\p)^4\e^2}
+{\l^3\fh^4\ov 8(4\p)^4\e}
-{b_1\l\fh^4\ov 4(4\p)^2\e}
+{c_1\l^2\fh^4\ov 4(4\p)^2\e}\nn\\
&+&{a_1\l^2\fh^2\ov (4\p)^2}\biggl[-{1\ov 4}+{\g\ov 4}+{1\ov 4}\ln\biggl(     
{\l\fh^2/2\ov 4\p M^2}\biggr)\biggr]
+{b_1\l\fh^4\ov (4\p)^2}\biggl[-{1\ov 8}+{\g\ov 8}+{1\ov 8}\ln\biggl(     
{\l\fh^2/2\ov 4\p M^2}\biggr)\biggr] \nn\\
&-&{c_1\l^2\fh^4\ov (4\p)^2}\biggl[-{1\ov 8}+{\g\ov 8}+{1\ov 8}\ln\biggl(     
{\l\fh^2/2\ov 4\p M^2}\biggr)\biggr] \nn\\
&+&{\l^3\fh^4\ov(4\p)^4}\biggl[{11\ov 32}+{A\ov 8}-{5\ov 16}\g 
+{3\ov 32}\g^2     
-{5\ov 16}\ln\biggl({\l\fh^2/2\ov 4\p M^2}\biggr)+{3\ov 16}\g     
\ln\biggl({\l\fh^2/2\ov 4\p M^2}\biggr)
+{3\ov 32}\ln^2\biggl({\l\fh^2/2\ov 4\p M^2}\biggr)   
\biggr]    
\;.\label{ve2}    
\eea 
In the above equation, $A$ is a constant whose value is defined
in Eq.~(\ref{AB}). Notice that the so-called ``dangerous'' pole terms such    
as $[\fh^l/\e^m]\ln^n [\l\fh^2/(4\p M^2)]$, $(l=0,2,4;$ $~m=1,2;~n=1,2)$,   
in the above equation, which    
cannot be removed by terms of counterterm constants ($\d m^2 \fh^2/2$    
and $\d\l\fh^4/(4!)$), have been completely     
cancelled out among each other. The counterterm constants $\d m_2^2$ and  
$\d\l_2$ are determined as 
\bea     
\d m_2^2&=&{a_1\l\ov (4\p)^2\e}+a_2\;,\nn\\     
\d\l_2&=&{9\l^3\ov (4\p)^4\e^2}
-{3\l^3\ov (4\p)^4\e}+{6b_1\l\ov (4\p)^2\e}
-{6c_1\l^2\ov (4\p)^2\e}+b_2\;, \label{22}
\eea    
where $a_2$ and $b_2$ are also unspecified but finite constants.

In the massive $O(N)$ $\f^4$ theory, the renormalization conditions 
\beas
\~{\G}^{(2)} (0)=-m^2\;,~~~~
\~{\G}^{(4)} (0)=-\l\;,
\eeas
are respectively translated into 
\beas
{d^2 V_{\rm eff}(\fh)\ov d\fh^2}\bigg |_{\fh=0}=m^2\;,~~~~   
{d^4 V_{\rm eff}(\fh)\ov d\fh^4}\bigg |_{\fh=0}=\l\;.   
\eeas 

In our massless theory, however, we encounter the infrared singularity in the 
defining condition for a coupling constant. To avoid this difficulty we 
follow Coleman and Weinberg \cite{cw} and require   
\bea   
{d^2 V_{\rm eff}(\fh)\ov d \fh^2}\bigg |_{\fh=0}=0\;,~~~~   
{d^4 V_{\rm eff}(\fh)\ov d \fh^4}\bigg |_{\fh^2=M^2}=\l\;.\label{rc}   
\eea   
Then constants $a_1$, $a_2$, $b_1$, and $b_2$ are determined order by order 
as follows:  
\bea  
a_1&=&a_2=0\;,\nn\\     
b_1&=&-{\l^2\ov (4\p)^2}\biggl[4+{3\ov 2}\g+{3\ov 2}
\ln\biggl({\l/2\ov 4\p}\biggr)
\biggr]\;,\nn\\    
b_2&=&{\l^3\ov (4\p)^4}\biggl[{139\ov 4}-3A+15\g+{9\ov 4}\g^2   
+\biggl(15+{9\ov 2}\g\biggr)\ln\biggl({\l/2\ov 4\p}\biggr)   
+{9\ov 4}\ln^2\biggl({\l/2\ov 4\p}\biggr)\biggr] \nn\\
&&+{c_1\l^2\ov (4\p)^2}\biggl[{19\ov 2}+3\g+3\ln\biggl({\l/2\ov
4\p}\biggr)\biggr]\;.\label{1122}  
\eea  
After disposing successfully all divergent terms in Eqs.~(\ref{ve1})
and (\ref{ve2}) by the counterterm constants in Eqs.~(\ref{11})
and (\ref{22}), we eventually arrive at our new result satisfying the 
conditions in Eq.~(\ref{rc}):
\bea      
V_{\rm eff}^{[2]}(\fh)&=&   
\biggl[{\l\ov 4!}\fh^4\biggr]   
+{\hbar\l^2\fh^4\ov (4\p)^2}\biggl[-{25\ov 96}+{1\ov 16}     
\ln\biggl({\fh^2\ov M^2}\biggr)     
\biggr]
+{\hbar^2\l^3\fh^4\ov (4\p)^4}\biggl[{55\ov 24}-{13\ov 16}     
\ln\biggl({\fh^2\ov M^2}\biggr)+{3\ov 32}     
\ln^2\biggl({\fh^2\ov M^2}\biggr)\biggr]\nn\\  
&+&\underline{{c_1\hbar^2\l^2\fh^4\ov (4\p)^2}\biggl[{25\ov 48}-{1\ov 8}     
\ln\biggl({\fh^2\ov M^2}\biggr)\biggr]}\;.\label{2p}    
\eea 
Our result differs from that of Jackiw (see Eq.~(3.17) in Ref.~\cite{jk})
just by the underlined square-bracket term in Eq.~(\ref{2p}) which is the 
missing portion in his calculation of the two-loop effective potential. 
After substituting
the value of $c_1$ of Eq.~(\ref{c1}), we have the final form of the 
effective potential up to the two-loop order as follows:
\bea      
V_{\rm eff}^{[2]}(\fh)&=&   
\biggl[{\l\ov 4!}\fh^4\biggr]   
+{\hbar\l^2\fh^4\ov (4\p)^2}\biggl[-{25\ov 96}+{1\ov 16}     
\ln\biggl({\fh^2\ov M^2}\biggr)     
\biggr]
+{\hbar^2\l^3\fh^4\ov (4\p)^4}\biggl[{635\ov 288}-{19\ov 24}     
\ln\biggl({\fh^2\ov M^2}\biggr)+{3\ov 32}     
\ln^2\biggl({\fh^2\ov M^2}\biggr)\biggr]\;.\label{f2p}    
\eea

%%%%%%%%%%%%%%%%%%%%%%%%%%%%%%%%%%%%%%%%%%%%%%%%%%%%%%%%%%%%%%%%%%%%%%%%%%%% 
\section{Discussion and Conclusion}       
%%%%%%%%%%%%%%%%%%%%%%%%%%%%%%%%%%%%%%%%%%%%%%%%%%%%%%%%%%%%%%%%%%%%%%%%%%%% 
Let us now apply a renormalization condition   
$d^4 V_{\rm eff}(\fh)/d \fh^4|_{\fh^2=M^2}=\l$ 
to the two-loop effective potential with most general form of 
$V_{\rm eff}^{(2)}$ as an 
assumed series solution to the renormalization group equation:  
\bea  
V_{\rm eff}^{[2]}(\fh)&=&\biggl[{\l\ov 4!}\fh^4\biggr]   
+{\hbar\l^2\fh^4\ov (4\p)^2}\biggl[-{25\ov 96}+{1\ov 16}     
\ln\biggl({\fh^2\ov M^2}\biggr)     
\biggr]
+{\hbar^2\l^3\fh^4\ov (4\p)^4}\biggl[
\a_0+\a_1\ln\biggl({\fh^2\ov M^2}\biggr)+  
\a_2\ln^2\biggl({\fh^2\ov M^2}\biggr)  
\biggr]\;,\label{im}  
\eea  
where $\a_0$, $\a_1$, $\a_2$ are constants. Then we readily obtain 
\bea  
{d^4 V_{\rm eff}^{[2]}(\fh)\ov d \fh^4}\bigg |_{\fh^2=M^2}&=&\l+ 
\fbox{${\displaystyle{\hbar^2\l^3\ov (4\p)^4}}\Bigl[24\a_0+100\a_1+280\a_2 
\Bigr]$}\;.\label{rrc} 
\eea  
The boxed term ($\l$-cubic term) on the right-hand side of the above equation 
is an unwanted term. Thus it should vanish. 
%Although this vanishing 
%boxed-cubic-term gives us $\a_0$ in terms of $\a_1$ and $\a_2$, this 
%information is not used in the following, because $\a_0$ disappears 
%automatically in a consistent way.
 
Next let us require the parametrization invariance of the theory. 
The renormalization mass, $M$, is indeed an arbitrary parameter,    
with no effect on the physics of the problem.    
If we pick a different mass, $M'$, then we define a new coupling constant   
\bea   
\l'={d^4V_{\rm eff}^{[2]}(\fh)\ov d\fh^4}\bigg|_{\fh^2=M'^{\,2}}=\l+   
P_1\l^2+P_2\l^3\;,\label{rp}   
\eea   
where   
\bea    
P_1&=&{3\hbar\ov 2(4\p)^2}\ln\biggl({M'^2\ov M^2}\biggr)\;,\nn\\   
P_2&=&{\hbar^2\ov (4\p)^4}\biggl[
%24\a_0+100\a_1+280\a_2+
(24\a_1+200\a_2)\ln\biggl({M'^{\,2}\ov M^2}\biggr)
+24\a_2\ln^2\biggl({M'^{\,2}\ov M^2}\biggr) 
\biggr]\;.\nn   
\eea 
Eq.~(\ref{rp}) is readily inverted 
%perturbatively 
iteratively as   
\beas   
\l=\l'-P_1\l'^{\,2}+(2P_1^2-P_2)\l'^{\,3}+O(\l'^{\,4})\;.   
\eeas
We now substitute this $\l$ into Eq.~(\ref{im}). Then the two-loop
effective potential is given in terms of $\l'$ and $M'$ as follows:
\bea     
V_{\rm eff}^{[2]}(\fh)&=&   
\biggl[{\l'\ov 4!}\fh^4\biggr]   
+{\hbar\l'^{\,2}\fh^4\ov (4\p)^2}\biggl[-{25\ov 96}+{1\ov 16}     
\ln\biggl({\fh^2\ov M'^{\,2}}\biggr)
\biggr]\nn\\
&&+{\hbar^2\l'^{\,3}\fh^4\ov (4\p)^4}\biggl[  
%-{25\ov 6}\a_1-{35\ov 3}\a_2
\a_0+\a_1\ln\biggl({\fh^2\ov M'^{\,2}}\biggr)+  
\a_2\ln^2\biggl({\fh^2\ov M'^{\,2}}\biggr)\nn\\ 
&&+\,\fbox{$\displaystyle{
\biggl\{{25\ov 32}-{25\ov 3}\a_2+\biggl(-{3\ov 16}+2\a_2\biggr) 
\ln\biggl({\fh^2\ov M'^{\,2}}\biggr)\biggr\} 
\ln\biggl({\displaystyle{M'^{\,2}\ov M^2}}\biggr)}$}~\biggr]
+O(\l'^{\,4})\;.\label{uw} 
\eea 
The parametrization invariance requires that 
the boxed term in Eq.~(\ref{uw}) should vanish. 
From this and the vanishing boxed term of Eq.~(\ref{rrc}) we obtain 
\bea 
V_{\rm eff}^{(2)}(\fh)&=&{\hbar^2\l^3\fh^4\ov (4\p)^4}\biggl[  
-{35\ov 32}-{25\ov 6}\a_1+\a_1\ln\biggl({\fh^2\ov M^2}\biggr)  
+{3\ov 32}\ln^2\biggl({\fh^2\ov M^2}\biggr)\biggr]\;.\label{2lp}  
\eea  
This shows us that even if one has an arbitrary 
value of $\a_1$, the parametrization invariance
still holds. This is the reason why the Jackiw's result (Eq.~(3.17) 
in Ref.~\cite{jk}) is safe from the check of the parametrization invariance.
In the above equation $\a_1$ is fixed not by the 
parametrization invariance but by the correct two-loop calculation of the 
effective potential. 

In summary, Jackiw used a wrong 
renormalization condtion, Eq.~(\ref{wzc}), 
in the massless $O(N)$ $\f^4$ theory and obtained such an incorrect value 
of $\a_1$ as $-{13\ov 16}$, but the 
correct value of $\a_1$ is $-{19\ov 24}$ as given by our Eq.~(\ref{f2p}).
%%%%%%%%%%%%%%%%%%%%%%%%%%%%%%%%%%%%%%%%%%%%%%%%%%%%%%%%%%%%%%%%%%%%%%%% 
\acknowledgments     
%%%%%%%%%%%%%%%%%%%%%%%%%%%%%%%%%%%%%%%%%%%%%%%%%%%%%%%%%%%%%%%%%%%%%%%% 
This work was supported in part by Ministry of Education, Project number     
BSRI-97-2442 and one of the authors (J.~-M. C.) was also supported in 
part by the Postdoctoral Fellowship of Kyung Hee University.

%%%%%%%%%%%%%%%%%%%%%%%%%%%%%%%%%%%%%%%%%%%%%%%%%%%%%%%%%%%%%%%%%%%%%%%%%%%%% 
\appendix  
%%%%%%%%%%%%%%%%%%%%%%%%%%%%%%%%%%%%%%%%%%%%%%%%%%%%%%%%%%%%%%%%%%%%%%%%%%%%% 
\section*{Loop Integrations}
\setcounter{equation}{0} 
%%%%%%%%%%%%%%%%%%%%%%%%%%%%%%%%%%%%%%%%%%%%%%%%%%%%%%%%%%%%%%%%%%%%%%%%%%%%%  
 
In this Appendix, the momenta appearing in the formulas are all    
(Wick-rotated) Euclidean ones and the abbreviated integration measure is   
defined as    
\beas    
\int_k=M^{4-n}\int{d^n k\ov (2\p)^n}\;,    
\eeas    
where $n=4-\e$ is the space-time dimension in the framework of dimensional   
regularization \cite{tv} and $M$ is an arbitrary constant with mass  
dimension.    
For the sake of completeness, we simply list one-loop and two-loop integrals  
needed in our calculations though they are well known. For the two-loop  
integrations one may refer to Ref.~\cite{2lp}. 
   
\subsection*{A. Loop integration formulas}        
\bea       
S_1&\equiv&\int_k\ln\biggl(1+{\xi^2\ov k^2+\s^2}\biggr)       
=-{(\xi^2+\s^2)^2\ov (4\p)^2}  
\biggl({\xi^2+\s^2\ov 4\p M^2}\biggr)^{\!\!\!-\e/2}\G\biggl({\e\ov 2}-2\biggr)  
+\,\xi\mbox{-independent term}\;,\nn\\   
S_2&\equiv&\int_k{1\ov k^2+\s^2}       
={\s^2\ov (4\p)^2}\biggl({\s^2\ov 4\p M^2}       
\biggr)^{\!\!\!-\e/2}\G\biggl({\e\ov 2}-1\biggr)\;,\nn\\       
S_3&\equiv&       
\int_k{1\ov (k^2+\s^2)^2}={1\ov (4\p)^2}\biggl({\s^2\ov 4\p M^2}       
\biggr)^{\!\!\!-\e/2}\G\biggl({\e\ov 2}\biggr)\;,\nn\\       
S_4&\equiv&\int_{k,\,p}{1\ov (k^2+\s^2)       
(p^2+\s^2)[(p+k)^2+\s^2]}
={\s^2\ov (4\p)^4}\biggl({\s^2\ov 4\p M^2}\biggr)^{\!\!\!-\e}       
{\G^2(1+\e/2)\ov (1-\e)(1-\e/2)}\biggl[-{6\ov \e^2}-3A+O(\e) 
\biggr]\;.\label{www}     
\eea       
In the above equation, $\g$ is the usual Euler constant,    
$\g=0.5772156649\cdots$, and the numerical value of the constant    
$A$ in Eq.~(\ref{www}) is       
\bea   
A=f(1,1)=-1.1719536193\cdots\;,\label{AB}
\eea     
where     
\beas     
&& f(a,b)\equiv\int_0^1dx\biggl[\int_0^{1-z}dy\biggl(-{\ln(1-y)\ov y} 
\biggr)       
-{z\ln z\ov 1-z}\biggr]\;,~~~z={ax+b(1-x)\ov x(1-x)}\;.     
\eeas        
\subsection*{B. Calculation of the diagrams in Fig.~3} 
\bea        
{\rm Diag.~1}&=&-{\hbar\ov 2(4\p)^2}\biggl({\d m^2+        
(\l+\d\l)\fh^2/2\ov 1+\d Z}\biggr)^{\!\!2}\biggl({\d m^2+(\l+\d\l)        
\fh^2/2\ov 4\p M^2(1+\d Z)}\biggr)^{\!\!\!-\e/2}\G\biggl(-2+ 
{\e\ov 2}\biggr)\nn\\        
&=&\hbar\biggl[-{\l^2\fh^4\ov 8(4\p)^2}        
\biggl({\l\fh^2/2\ov 4\p M^2}\biggr)^{\!\!\!-\e/2}        
\G\biggl(-2+{\e\ov 2}\biggr)\biggr]\nn\\        
&+&\hbar^2\biggl[{\l\ov (4\p)^2}\biggl(-{\d m_1^2\fh^2\ov 2} 
-{\d\l_1\fh^4\ov 4}+{\l\d Z_1\fh^4\ov 4}\biggr)        
\biggl(1-{\e\ov 4}\biggr)\biggl({\l\fh^2/2\ov 4\p M^2} 
\biggr)^{\!\!\!-\e/2}\G\biggl(-2+{\e\ov 2}\biggr)\biggr]\;,\nn\\        
\mbox{Diag.~2}       
&=&\hbar^2\biggl[{\l^3\fh^4\ov 32(4\p)^4} 
\biggl({\l\fh^2/2\ov 4\p M^2}        
\biggr)^{\!\!\!-\e}\G^2\biggl(-1+{\e\ov 2}\biggr)\biggr]\;,\nn\\        
\mbox{Diag.~3}
&=&\hbar^2\biggl[-{\l^3\fh^4\ov 24(4\p)^4} 
\biggl({\l\fh^2/2\ov 4\p M^2}        
\biggr)^{\!\!\!-\e}{\G^2(1+\e/2)\ov (1-\e)(1-\e/2)}\biggl(-{6\ov \e^2}        
-3A\biggr)\biggr]\;.\label{int}   
\eea     

%%%%%%%%%%%%%%%%%%%%%%%%%%%%%%%%%%%%%%%%%%%%%%%%%%%%%%%%%%%%%%%%%%%%%%% 
 
\end{document}